\documentstyle[12pt,epsfig]{article}

\newcommand{\la}{\lambda}

\newcommand{\tr}{\mbox{Tr}}
\newcommand{\Tr}{\mbox{\bf Tr}}
                         
\let\vf=\vfil \let\cl=\centerline                    

\let\a=\alpha \let\be=\beta \let\g=\gamma                         
\let\e=\varepsilon  \let\h=\eta                       
 \let\k=\kappa \let\la=\lambda \let\m=\mu                      
\let\n=\nu  \let\p=\pi \let\r=\rho \let\s=\sigma                      
                                                      
\let\ph=\varphi                             
\let\Om=\Omega                                      
 \let\G=\Gamma \let\D=\Delta                                     
                                                                                
\def\0{\over } \def\1{\vec }     \def\2{{1\over2}} \def\4{{1\over4}}            
\def\5{\bar }  \def\6{\partial } \def\7#1{{#1}\llap{/}}                         
\def\8#1{{\textstyle{#1}}}       \def\9#1{{\bf {#1}}}                           
 \def\llp{\hbox to 0pt{\hss /\hskip1.5pt}}    
\def\llo{\hbox to 0.2pt{\hss /}} \def\llq{\hbox to 0pt{\hss /\hskip0.5pt}}      
\def\so{\supset\hbox to 0pt{\hss $\displaystyle -$\hskip1pt}}

\def\<{\langle } \def\>{\rangle }

\let\ap=\approx

\def\bea{\begin{eqnarray}} \def\eea{\end{eqnarray}}                             
\def\beann{\begin{eqnarray*}} \def\eeann{\end{eqnarray*}}                       
\def\beq{\begin{equation}} \def\eeq{\end{equation}}

\def\a{\alpha}
\def\b{\beta}

\def\d{\delta}
\def\e{\epsilon}
\def\f{\phi}
\def\g{\gamma}
\def\h{\eta}

\def\k{\kappa}
\def\l{\lambda}
\def\m{\mu}
\def\n{\nu}

\def\p{\pi}

\def\r{\rho}
\def\s{\sigma}
\def\t{\tau}

\def\D{\Delta}
\def\F{\Phi}
\def\G{\Gamma}

\def\Q{\Theta}

\def\vf{\varphi}

\def\cl{{\cal L}}

\def\co{{\cal O}}

\def\bo{{\raise.15ex\hbox{\large$\Box$}}}               
\def\pa{\partial}                                       
\def\pr{\prod}                                          
\def\face{{\raise.2ex\hbox{$\displaystyle \bigodot$}\mskip-2.2mu \llap {$\ddot
        \smile$}}}                                      
\def\dg{\dagger}                                     

\def\VEV#1{\left\langle #1\right\rangle}        
\def\leftrightarrowfill{$\mathsurround=0pt \mathord\leftarrow \mkern-6mu
        \cleaders\hbox{$\mkern-2mu \mathord- \mkern-2mu$}\hfill
        \mkern-6mu \mathord\rightarrow$}       
\def\dvec#1{\vbox{\ialign{##\crcr
        \leftrightarrowfill\crcr\noalign{\kern-1pt\nointerlineskip}
        $\hfil\displaystyle{#1}\hfil$\crcr}}}           

\def\beq{\begin{equation}}
\def\eeq{\end{equation}}

\def\beqx{\begin{displaymath}}
\def\eeqx{\end{displaymath}}

\def\beqa{\begin{eqnarray}}
\def\eeqa{\end{eqnarray}}
\def\NO{\nonumber}

\def\pl#1#2#3{Phys.~Lett.~{B{#1}} (19{#2}) #3}
\def\np#1#2#3{Nucl.~Phys.~{B{#1}} (19{#2}) #3}
\def\prl#1#2#3{Phys.~Rev.~Lett.~{#1} (19{#2}) #3}
\def\pr#1#2#3{Phys.~Rev.~{D{#1}} (19{#2}) #3}

\def\ap#1#2#3{Ann.~of Phys.~{{#1}} (19{#2}) #3}

\def\mpl#1#2#3{Mod.~Phys.~Lett.~{A{#1}} (19{#2}) #3}
\def\nc#1#2#3{Nuovo Cim.~{{#1}} (19{#2}) #3}

\date{}
\title{
{\large\rm DESY 96-216}\hfill{\large\tt ISSN 0418-9833}\\
{\large\rm October 1996}\hfill\vspace*{2cm}\\
Electroweak Phase Transition\\ 
and Neutrino Masses\footnote{\noindent To be published in {\it Quarks '96}, 
Proc. of the IXth International Seminar, Yaroslavl, Russia, 1996}}
\author{W. Buchm\"uller \\
\vspace{5.0\baselineskip}                                               
{\normalsize\it Deutsches Elektronen-Synchrotron DESY, 22603 Hamburg, Germany}
\vspace*{2cm}\\                     
}                                                                          
\begin{document}                                                  

\maketitle  
\begin{abstract}
\noindent
The presently observed cosmological baryon asymmetry has been finally
determined at the time of the electroweak phase transition, when baryon 
and lepton number violating interactions fell out of thermal equilibrium. 
We discuss the thermodynamics of the phase transition based on the free
energy of the SU(2) Higgs model at finite temperature,
which has been studied in perturbation theory and lattice simulations.
The results suggest that the baryon asymmetry has been generated by lepton
number violating interactions in the symmetric phase of the standard model,
i.e., at temperatures above the critical temperature of the electroweak
transition. The observed value of the baryon asymmetry, $n_B/s \sim 10^{-10}$,
is naturally obtained in an extension of the standard model with 
right-handed neutrinos where $B-L$ is broken at the unification scale 
$\Lambda_{\mbox{\scriptsize GUT}}\sim 10^{16}$ GeV. The corresponding
pattern of masses and mixings of the light neutrinos $\n_e$, $\n_\m$ and
$\n_\t$ is briefly described.
\end{abstract} 
\thispagestyle{empty}

\newpage

\section{Introduction}

In the standard model of electroweak interactions all masses are generated by
the Higgs mechanism. As a consequence, at high temperatures a transition
occurs from a massive low-temperature phase to a `massless' high-temperature 
phase, where the Higgs vacuum expectation value `evaporates' and the 
electroweak symmetry is `restored' \cite{kirzh}. 

Due to the chiral nature of the weak interactions baryon number ($B$) and 
lepton number ($L$) are not conserved in the standard model \cite{thoo}. At 
zero temperature this has no observable effect due to the smallness of the weak 
coupling. However, as the temperature approaches the critical 
temperature $T_c$ of the electroweak phase transition, $B$ and $L$
violating processes come into thermal equilibrium \cite{krs}. Their rate 
is determined by the free energy of sphaleron-type field configurations 
which carry topological charge. In the standard model they induce
an effective interaction of all left-handed fermions (cf. Fig.~1)
which violates baryon and lepton number by three units,
\beq
\D B = \D L = 3\,  .
\eeq
\begin{figure}[bh]
\begin{center}
\epsfig{file=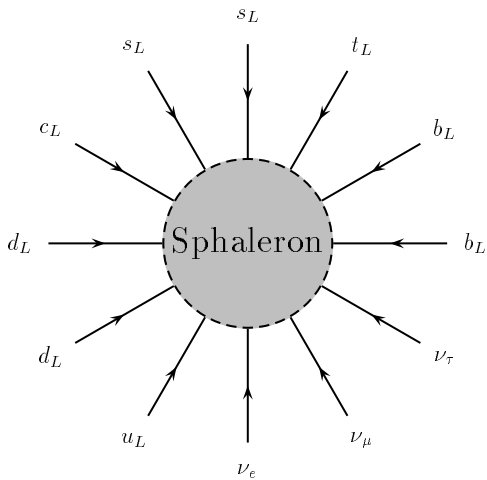,width=7cm}
\caption{One of the 12-fermion processes which are in thermal equilibrium
in the high-temperature phase of the standard model.}
\end{center}
\end{figure}
Since $B$ and $L$ violating processes fall out of thermal equilibrium below 
$T_c$, the presently observed value of the baryon asymmetry of the universe 
has finally been determined at the electroweak transition. Hence, the study
of the thermodynamics of this transition is of great cosmological significance.

Sphaleron processes conserve $B-L$. In the high-temperature, `symmetric' phase,
where $B$ and $L$ violating processes are expected to stay in thermal 
equilibrium over a wide range of temperatures, the expectation values of 
$B$ and $L$ are related,
\beq \label{bleq}
\langle B \rangle_T \simeq C \langle B-L \rangle_T 
\simeq {C\over C-1} \langle L \rangle_T \, ,
\eeq
where $C={28\over 79}$ in the standard model \cite{sphal}. Hence, baryon and
lepton number are correlated in the symmetric phase, and the generation of a
baryon asymmetry requires lepton number violation.

\begin{figure}[th]
\begin{center}
\epsfig{file=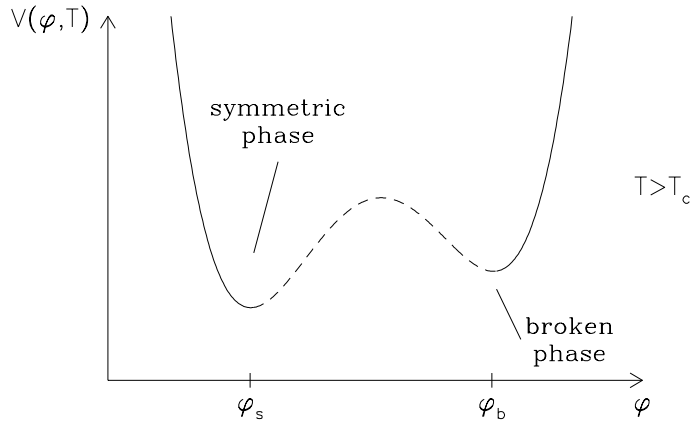,width=10cm}
\caption{Free energy as function of the order parameter at a temperature
slightly above the critical temperature of a first-order phase transition.}
\end{center}
\end{figure}

In the abelian Higgs model, i.e., scalar electrodynamics, the high-temperature
phase transition is known to be of first order for sufficiently small Higgs
masses \cite{halp,kili}. The expectation value of the Higgs field plays the
role of an order parameter. At temperatures just above the critical temperature
$T_c$ the Higgs phase is metastable. It decays in a first-order transition
to the symmetric phase, which is accompanied by a jump in the order parameter
$\D v = v_b - v_s$ (cf. Fig.~2).

A similar behaviour is expected in the standard model, with a critical 
temperature given by the Fermi constant, $T_c \sim G_F^{-1/2}$. In a 
first-order phase transition a strong deviation from thermal equilibrium
can occur. Since the standard model contains $CP$ violating and, at high
temperatures, also $B$ violating interactions, all conditions for baryogenesis
are fulfilled. It is therefore conceivable that the cosmological baryon
asymmetry has indeed been generated at the electroweak phase transition
\cite{rev}. Such a scenario requires that a produced baryon asymmetry is not 
erased by the sphaleron processes in the Higgs phase close to the critical 
temperature. From this condition a lower bound on the jump of order parameter
at the phase transition can be derived (see \cite{rev}),
\pagebreak
\beq\label{spcon}
\frac{\D v (T_c)}{T_c} > 1.2\, .
\eeq
This condition is necessary for electroweak baryogenesis although far from 
sufficient. A discussion of the complicated nonequilibrium processes in the 
electroweak plasma can be found in \cite{rev}.

In order to examine whether condition (\ref{spcon}) is satisfied, one
has to study the thermodynamics of the electroweak transition near the
critical temperature. Here the main obstacle are the well known infrared
divergencies of non-abelian gauge theories at high temperature. As a first
step towards a treatment of the full standard model the following 
simplifications are usually made: fermions are integrated out using 
high-temperature perturbation theory, and the electromagnetic interaction is
neglected ($\sin{\Q_W}=0$). One is then left with the SU(2) Higgs model.

\section{Thermodynamics of the SU(2) Higgs model}

The observables which characterize a first-order phase transition are critical 
temperature, jump in the order parameter and latent heat. Also important are
surface tension and correlation lengths which are more difficult to compute
and which we shall not discuss in detail. Our analysis will be based on the
gauge invariant `order parameter' $\langle\F^{\dagger}\F\rangle$ and the 
corresponding free energy computed in lattice simulations and perturbation 
theory. In our discussion we shall closely follow \cite{bfh}. 

It is well known that the electroweak phase transition is influenced by
non-perturbative effects whose size is governed by the confinement scale of the
effective three-dimensional theory which describes the high-temperature limit
of the SU(2) Higgs model. These effects are particularly relevant in the
symmetric phase, and one may worry to what extent a perturbative analysis of 
the phase transition can yield sensible results. As we shall see, perturbation
theory is self-consistent at two-loop order. A comparison with 
results obtained by lattice simulations will then enable us to estimate the 
effect of non-perturbative corrections.\\

{\noindent\it Generalities}\\

The action of the SU(2) Higgs model at finite temperature $T$ reads
\beq \label{ST}
S_{\beta}[\F,W] = \int_{\beta} dx \; \tr \left[
{1\over 2}W_{\mu\nu}W_{\mu\nu} + 
(D_{\mu}\F)^\dg D_{\mu}\F + \mu \F^{\dagger} \F 
+ 2 \lambda (\F^{\dg} \F)^2 \right] \, , 
\eeq
with 
\beqa
\F &=& \2 (\s + i \vec{\pi}\cdot \vec{\tau}) \, ,\quad 
D_{\mu}\F = (\6_{\mu} - i g W_{\mu})\F\, ,\quad  
W_{\mu} = \2\vec{\tau}\cdot \vec{W_{\mu}}\ ,\\
&&\quad \int_{\beta}dx = \int_0^{\beta}d\tau\int_{\Om}d^3x\ ,\quad 
\beta = {1\over T}\ .
\eeqa
Here $\vec{W_{\mu}}$ is the vector field, $\s$ is the Higgs field, $\vec{\pi}$
is the Goldstone field, $\vec{\tau}$ is the triplet of Pauli matrices,
and $\Om$ is the spatial volume. For perturbative calculations gauge fixing 
and ghost terms have to be added to the action (\ref{ST}).

The free energy density of the system, $W(T,J)$, is given by the partition
function, i.e., the trace of the density matrix,
\beq 
\exp(-\be\Om W(T,J)) = \Tr \exp\left[-\be\left(\hat{H}+
J \int_{\Om}d^3x \hat{\F}^{\dg}\hat{\F}\right)\right]\ ,
\eeq
where $\hat{H}$ is the Hamilton operator of the theory, and $\hat{\F}$ is the
operator describing the Higgs field. We have added a source $J$, with
$\6_{\mu}J = 0$, coupled to the spatial average of the gauge invariant
composite operator $\hat{\F}^{\dg}\hat{\F}$ (here and below the trace
operator acting on $\hat{\F}^{\dg}\hat{\F}$ is omitted for brevity).
The partition function can be expressed as a euclidian functional integral
(see \cite{kapusta}),
\beq \label{FI}
\exp\left(-\be\Om W(T,J)\right) = \int_{\be}D\F D\F^{\dg} DW_{\mu}
\exp\left(-\int_{\be}dx\left(L + J \F^{\dg}\F\right)\right)\ ,
\eeq
where {\it L} is the euclidean lagrangian density, and
the bosonic fields $\F$ and $W_{\mu}$ satisfy periodic boundary
conditions at $\tau =0$ and $\tau = \be$. Eq. (\ref{FI}) is the
starting point of perturbative as well as numerical evaluations of the free
energy.

Note, that the source $J$ in Eq.~(\ref{FI}) couples to a gauge invariant
composite field.  Hence, the free energy $W(T,J)$ is gauge independent.
The spatially constant source $J$ simply redefines the mass term in the
action (\ref{ST}). This is in contrast to the usually considered generating 
function of connected Green functions at zero momentum,
\beq
\exp\left(-\be\Om \tilde{W}(T,j;J)\right) = 
\int_{\be}D\F D\F^{\dg} DW_{\mu}\exp\left(-\int_{\be}dx\left(L + 
J \F^{\dg}\F + j\s \right)\right)\ .
\eeq
Here the source $j$ couples to a gauge dependent quantity, the field $\s$.
Consequently, $\tilde{W}(T,j;J)$ is gauge dependent and not a
physical observable. We have also kept the dependence on the source $J$.
From $\tilde{W}(T,j;J)$ one obtains the effective potential $\tilde{V}
(T,\ph;J)$ via Legendre transformation, with $\ph={\pa \tilde{W}/ \pa j}$. 
The wanted free energy density $W(T,J)$ can now be obtained
from the effective potential $\tilde{V}$. In the infinite volume limit, 
one has
\beq\label{conn}
W(T,J) = \tilde{V}(T,\ph_{min}(T,J),J)\ ,
\eeq
where $\ph_{min}(T,J)$ is the global minimum of the effective potential
$\tilde{V}(T,\ph;J)$. For arbitrary values of $\ph$ the potential $\tilde{V}$
is gauge dependent. However, its value at the minimum is known to be
gauge independent, yielding a gauge independent free energy
$W(T,J)$.

At the critical temperature $T_c$ of a first-order transition the order 
parameter $\rho$,
\beq
\2 \rho \equiv {1\over \Om}\int_{\Om}d^3x \langle\hat{\F}^{\dg}(x)\hat{\F}(x)
\rangle = {\6\over \6 J}W(T,J)\ , 
\eeq
and the energy density $E$,
\beq
E(T,J) = W(T,J) - T {\6\over \6 T}W(T,J)\ ,
\eeq
are discontinuous. The jump in the energy density is the latent heat
$\Delta Q$. 

For the first-order phase transition from liquid to vapour there exists
a well known relation between the latent heat and the change of the
molar volume, the Clausius-Clapeyron equation \cite{callan}. In the
electroweak phase transition the order parameter 
$\langle \F^{\dg}\F \rangle$
plays the role of the molar volume, and a completely analogous relation
can be derived.

The electroweak plasma can exist in two phases, the massive low-temperature
Higgs phase with free energy $W_b(T,J)$ and the massless 
high-temperature symmetric phase with free energy $W_s(T,J)$. In the
$J-T$-plane the boundary between the two phases is determined by the
equilibrium condition
\beq \label{equi}
W_s(T,J(T)) = W_b(T,J(T))\ .
\eeq
This equilibrium condition yields a useful connection between the latent heat 
$\D Q$ and the jump in the order parameter $\D \rho$. Using the definitions
\beq
\D Q = - T {\6\over \6 T}\left(W_s - W_b\right) \quad,\quad
\D \rho = 2\ {\6\over \6 J}\left(W_s - W_b\right)\, ,
\eeq
one easily obtains
\beq\label{cceq}
\D Q = \2 \D\rho\, T{d J\over d T}\ .
\eeq
This is the Clausius-Clapeyron equation of the electroweak phase transition.
Together with a dimensional analysis it implies
\beq
\D Q = - \2 m_H^2 \D \rho (1 + {\cal O}(g^2, \la))\ ,
\label{cc}\eeq
where $m_H = \sqrt{-2\m^2}(1 + \co(g^2,\l))$ is the physical Higgs mass at
zero temperature. This relation provides a useful check for perturbative as
well as lattice results.\\

{\noindent\it Perturbation theory}\\

Near the ground state, $J = 0$, the free energy $W(T,J)$ can be evaluated 
as power series in the couplings $g$ and $\la$ by means of resummed 
perturbation theory which has been carried out up to two loops 
\cite{arnold,hebecker,bfh}. Here, thermal corrections are added to the 
tree-level masses of the scalar fields and the longitudinal component of 
the vector boson field, 
\beq \label{CT}
\delta S_{\be} = \be \int d^3x \left(\2 \a_{01} T^2 (\s^2 + \pi^2)
+ \2 \a_1 T^2 W_L^2\right)\ .
\eeq
The sum of tree-level masses and thermal corrections then enters the boson
propagators in loop diagrams, and $\d S_{\be}^c = -\d S_{\be}$ is treated as
counter term. It turns out that the resummation of static modes only is a
preferred procedure \cite{arnold}. Hence, in Eq.~(\ref{CT}) the fields $\s$ 
and $\pi$ do not depend on the imaginary time $\tau$. To leading order in 
the couplings, one obtains for the parameters in Eq.~(\ref{CT}) from 
one-loop self energy corrections $\a_{01} = {3\over 16}g^2 + \2 \la$,
$\a_1 = {5\over 6}g^2$.

The masses of the boson propagators are obtained from Eqs.~(\ref{ST}) and
(\ref{CT}) by shifting the Higgs field $\s$ by the average field
$\ph$. This yields $m_L$, $m_T$, $m_{\s}$ and $m_{\pi}$ for longitudinal 
and transverse part of the vector field, the Higgs field and the Goldstone 
field, respectively. The resummation procedure can be optimized by adding terms
of higher order in the couplings to Eq.~(\ref{CT}). In the Higgs phase,
where one is only interested in the effective potential close to the
minimum, the choice of a field dependent correction $\d S^b_\b$ is useful
which yields for the scalar masses,
\beq\label{msb}
m_{\s}^2 = 2\la \ph^2 \quad ,\qquad m_{\pi}^2 = 0\ .
\eeq
Hence, no thermal resummation for scalar masses is performed in the Higgs
phase. In the symmetric phase it is useful to determine the scalar masses
self-consistently by 
\beq\label{selfc}
m_{\s}^2 = m_{\pi}^2 = {1\over \ph}{\6\over \6\ph}\tilde{V}(T,0;J)\ .
\eeq
For given vector boson masses $m_L$ and $m_T$, this is a gap equation
for the scalar masses, which can be solved at each order of the loop
expansion.

Having specified the vector boson and scalar masses in the Higgs phase and in 
the symmetric phase, the free energy can be calculated from the effective 
potential $\tilde{V}$, using the relation (\ref{conn}). Its two local
minima yield the free energy $W_s(T,J)$ and $W_b(T,J)$ in the symmetric
phase and the Higgs phase, respectively. The free energy of the ground state is
\beq
W(T,0) = \min\{W_s(T,0), W_b(T,0)\}\ .
\eeq
It is a concave function whose derivative is discontinuous at the critical 
temperature $T_c$. We can also consider the dependence of the free energy
on the external source at the critical temperature $T_c$. This function,
$W(T_c,J)$, is shown in Fig.~3. Here we have subtracted in both phases
the huge linear term $T^2J/6$, whose sole effect is to shift the expectation 
value $\langle\F^{\dagger}\F\rangle$ by $T^2/3$.

From the free energy $W(T,J)$ one can obtain the gauge invariant
effective potential $V(T,\rho)$ by means of a Legendre transformation.
Since the derivative of $W(T,J)$ is not continuous everywhere, one
has to use the definition (see \cite{oraef}),
\beq
V(T,\rho) = \sup_J\{W(T,J) - \2 \rho J\}\ .
\eeq
\begin{center}
\epsfig{file=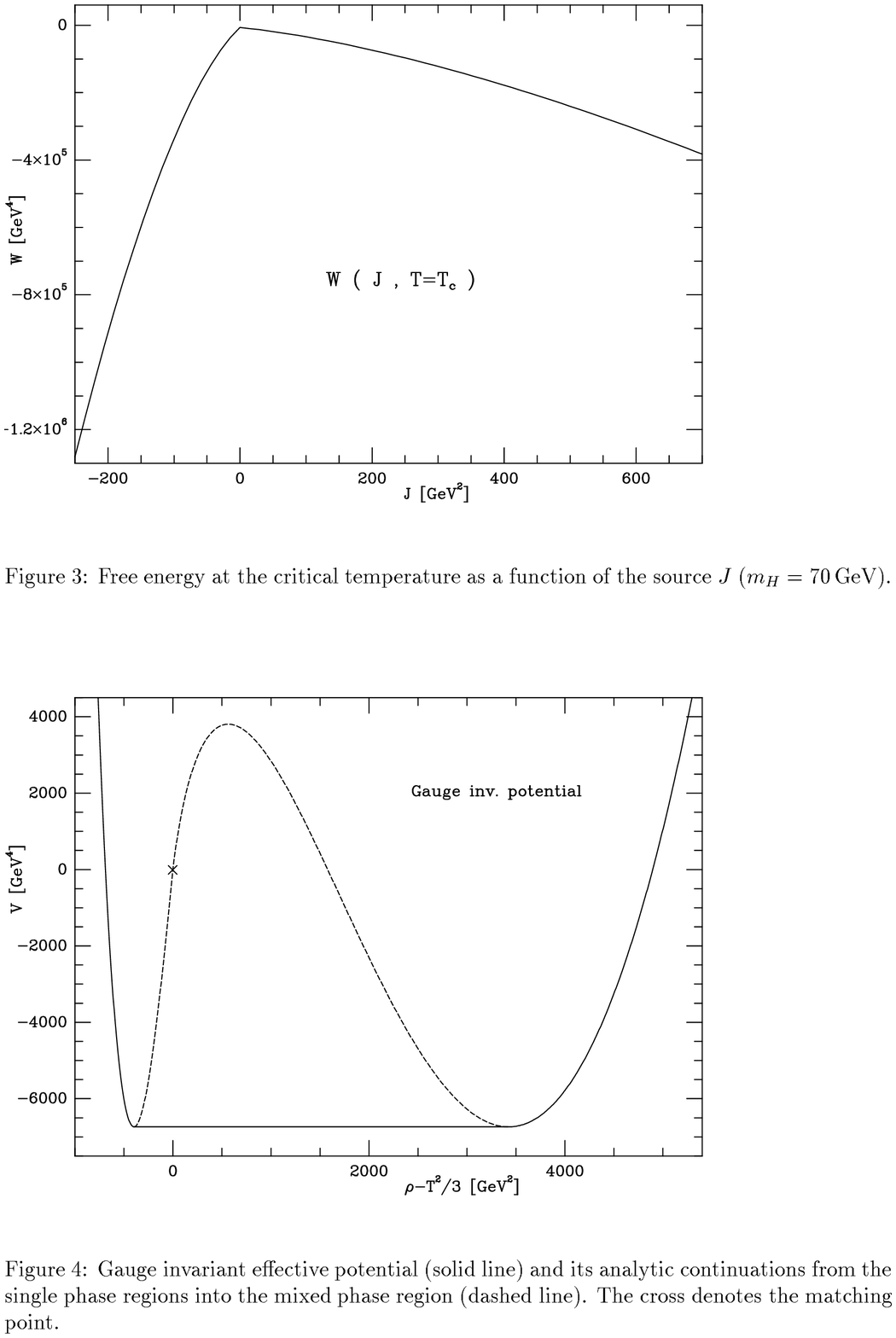,bbllx=50pt,bblly=39pt,bburx=546pt,bbury=788pt,width=14cm}
\end{center}
\clearpage
\begin{center}
\epsfig{file=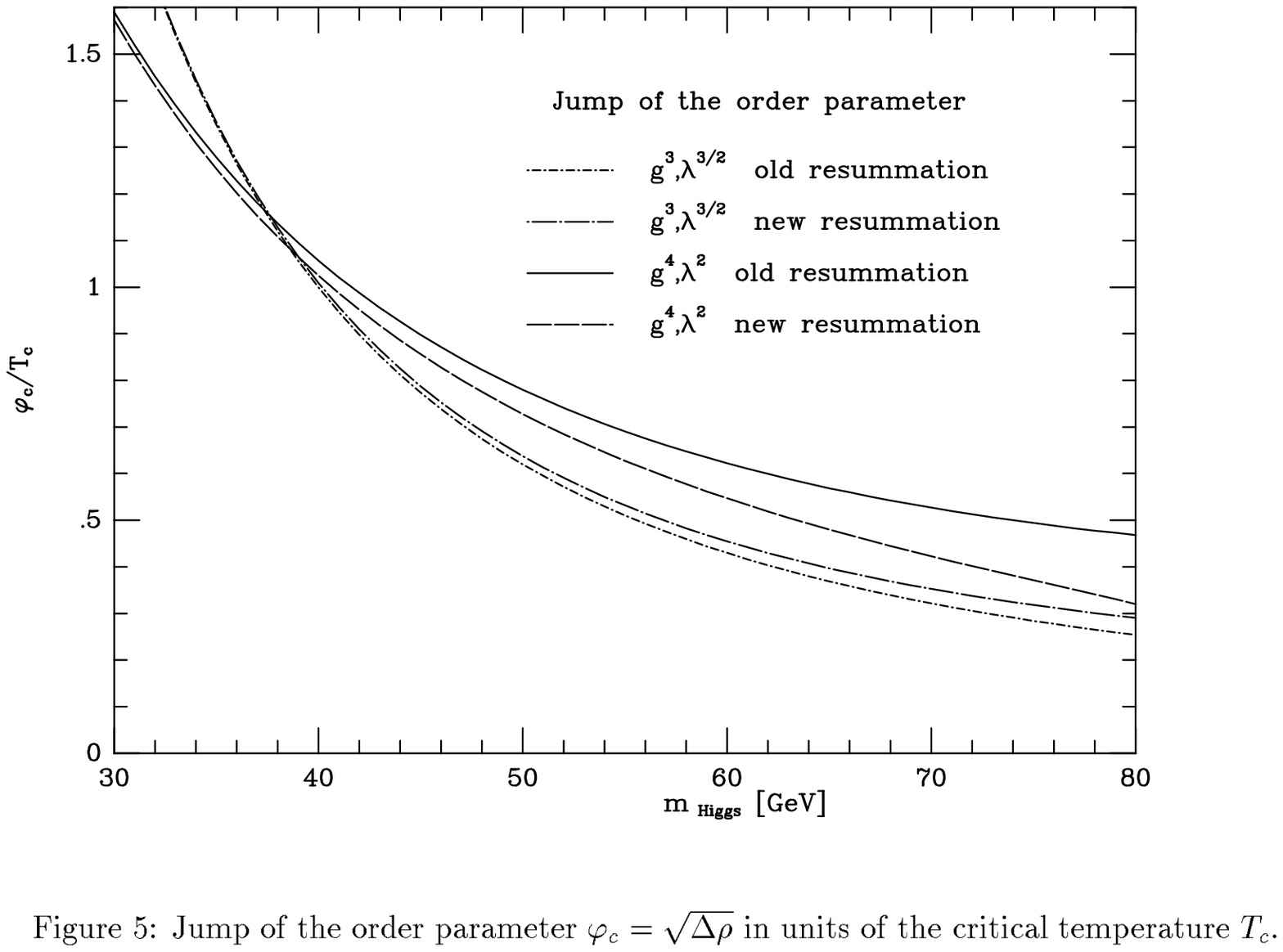,bbllx=62pt,bblly=229pt,bburx=529pt,bbury=594pt,width=14cm}
\end{center}
This yields the convex, non-analytic function, which is plotted in Fig.~4 as
full line. One may also compute the ordinary Legendre transform
$V_s(T,\rho)$ and $V_b(T,\rho)$ of $W_s(T,J)$ and $W_b(T,J)$, respectively. 
$V_s$ and $V_b$ are also shown in Fig.~4. In the region outside of the two
local minima, $V_s$ and $V_b$, respectively, coincide with the convex effective
potential $V(T,\rho)$. Between the two local minima, $V_s$ and $V_b$ represent
two analytical continuations of $V(T,\rho)$, which meet at the `matching
point' $\rho_M = T^2/3$. At this point, marked by a cross in the plot, the
first derivatives of both curves coincide.

The non-convex `effective potential' obtained by combining $V_s$ and $V_b$ on
both sides of the `matching point' has a barrier between symmetric and Higgs
phase like the ordinary effective potential $\tilde{V}(T,\ph;J)$. Note,
however, the difference in the range of fields. For $\tilde{V}$ one has
$0\leq \ph < \infty$, whereas for the potential $V$ the field $\rho$ varies
in the range $-\infty < \rho < \infty$. The generation of a barrier between two
local minima as analytic continuation from a convex effective potential is
reminiscent of the treatment of first-order phase transitions in condensed
matter physics \cite{langer}. However, the precise physical meaning of the
resulting non-convex `effective potential' still remains to be understood.\\

\pagebreak
{\noindent\it Comparison with lattice simulations}\\

The asymmetric resummation described above has been carried out up to 
two-loop order. In the Higgs phase the scalar masses are given by 
Eq.~(\ref{msb}), and in the symmetric phase they are self-consistently 
determined from 
Eq.~(\ref{selfc}). Note, that in the two-loop calculation the counter term 
to be inserted in the one-loop graph is ${\cal O}(g^3)$, whereas the scalar 
mass determined from Eq.~(\ref{selfc}) is of higher order in $g$.

In the symmetric phase the self-consistently determined scalar masses are
infrared divergent. With $m_T = g\ph/2$, the two-loop potential yields 
a contribution which diverges logarithmically at $\ph \approx 0$, 
\beq\label{ldiv}
m^2_{\s} \simeq - {33 g^4\over 128\pi^2}T^2 \ln{\be m_T}\ .
\eeq
Following \cite{bfhw} one may regularize this divergence by means of a
`magnetic mass' term. In Eq.~(\ref{ldiv}) one substitutes
$m_T^2 = g^2\ph^2/4 + \gamma^2g^4 T^2/(9\pi^2)$. In the following numerical 
results will be given for $\gamma =1$, which is obtained by one-loop gap 
equations \cite{bfhw,espinosa}. The results change only insignificantly if 
the parameter $\gamma$ is varied between 0.3 and 3.0. In addition to the
resummation procedure one has to choose a renormalization scheme.  A good
choice is the $\overline{\mbox{MS}}$-scheme with $\bar{\mu}=T$, supplemented
by finite counter terms $\d \m^2$ and $\d \la$ which account for the most 
important zero-temperature renormalization effects. 

Given the two-loop potential in the symmetric phase and in the Higgs phase,
one can numerically determine the critical temperature $T_c$, where
the two potentials at their respective local minima are degenerate. 
Differentiation with respect to temperature and the external source $J$
then yields latent heat and jump in the order parameter 
$\rho = 2\langle\F^{\dg}\F\rangle$.
The result for $\ph_c = \sqrt{\D\rho}$ is shown in Fig.~5, labelled 
`new resummation'. Here the zero-temperature standard model values 
$m_W=$ 80 GeV and $g^2 = 0.57$ have been used.
In the case of the `old resummation' \cite{hebecker} $\varphi_c$ corresponds 
to the position of the second minimum. The one-loop potential is 
$\co (g^3,\la^{3/2})$, the two-loop potential is $\co (g^4,\la^2)$. As Fig.~5 
illustrates, the 
`new resummation' procedure improves the convergence significantly. The 
relative change of an observable from one-loop to two-loop may be 
characterized by $\d = 2 |O_1 - O_2|/(O_1 + O_2)$. For the jump in the
order parameter $\ph_c$, $\d$ increases from $\sim 0.01$ at $m_H = 40$
GeV to $\sim 0.2$ at $m_H = 70$ GeV. Above $m_H \sim 80$ GeV the convergence
deteriorates rapidly, and the perturbative calculation is no longer
self-consistent.

Observables of the four-dimensional SU(2) Higgs model at finite temperature
can be directly computed by means of Monte Carlo simulations on lattices 
with spatial size $L_s$ and temporal size $L_t \ll L_s$. Such finite
temperature simulations at small values of $\la$ were initiated in \cite{bunk},
and high statistics  simulations were performed in \cite{montvay,csikor}. 
For the two Higgs masses $m_H \simeq 18$ GeV and $m_H \simeq 49$ GeV the
quantities $T_c$, $\D Q$ and $\ph_c \equiv v_T$ were computed on $L_t =2$
and $L_t =3$ lattices, and for $m_H \simeq 35$ GeV a detailed study of the
first-order phase transition was carried out on lattices of temporal size
up to $L_t = 5$.

In a complementary approach detailed studies have also been carried out  
based on dimensional reduction where non-zero Matsubara frequencies 
are first integrated out perturbatively. Recent results on dimensional 
reduction are described in \cite{mack,jako} where also references to previous 
work can be found.
Numerical simulations were performed for the effective three dimensional 
theory which led to a quantitative description of the first-order phase 
transition for Higgs masses up to $\sim 70$ GeV \cite{kaj,gurt}. 

\begin{figure}[htb] 
\setcounter{figure}{5}
\centerline{ \epsfysize=11.0cm
             \epsfxsize=9.57cm
             \epsfbox{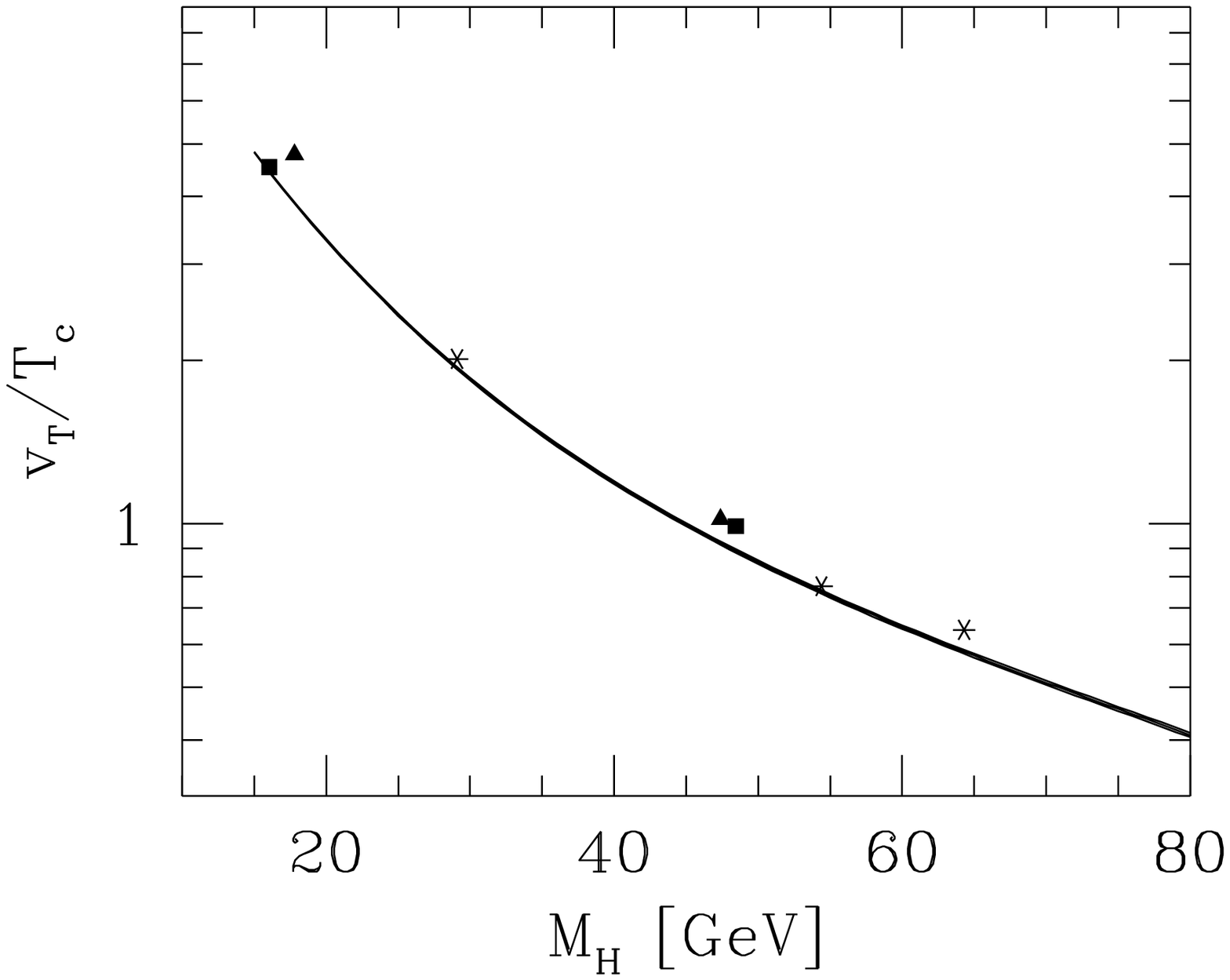}}
\vspace{-25mm}
\caption{Jump of the order parameter at the critical temperature. Comparison
of four-dimensional simulations (triangles, squares 
\protect{\cite{montvay}}) with three-dimensional simulations (stars 
\protect{\cite{kaj}}) and perturbation theory \protect{\cite{bfh}}. 
From \protect{\cite{jan}}.}
\end{figure}

In Fig.~6 results for the jump in the order parameter, 
$\D v(T_c)\equiv v_T$, are
compared \cite{jan}, which were obtained by simulations of the four-dimensional
theory \cite{montvay}, the three-dimensional theory \cite{kaj} and perturbation
theory \cite{bfh}, respectively. The comparison is made for the parameter
values $M_W = 80$ GeV and $g = 0.57$. The statistical errors of the
numerical simulations are so small that they are invisible in the figure.
The agreement between the three approaches is remarkable and certainly better 
than the systematic uncertainties of perturbation theory (cf.~Fig.~5). 
The continuum limit for the critical temperature has been studied in detail 
in \cite{csikor} for $m_H \simeq 34$ GeV , i.e., 
$R_{HW} = M_H/M_W \simeq 0.42$. The results are shown in Fig.~7, where the
error bars include the statistical error and an estimate of the systematic
error. A comparison is made with perturbation theory
and the three-dimensional simulation. The lattice simulations yield a
critical temperature slightly below the result from perturbation theory,
but the difference is not significant.

The agreement between results from perturbation theory and non-perturbative
lattice simulations is surprizing, since in the symmetric phase perturbation
theory is known to be infrared divergent, which prevents 
a straightforward extension of the
present two-loop calculation to three loops. The infrared behaviour of a
running gauge coupling has been studied in \cite{reuter}. In the symmetric 
phase the scalar masses depend logarithmically on an infrared cutoff of order 
the magnetic scale $\sim g^2 T$. This is reminiscent of the Debye screening
length in pure gauge theory which, at two-loop order, also requires an
infrared cutoff $\sim g^2 T$ \cite{reb}. Here, a non-perturbative definition
of the Debye screening length can be given such that the dominant contribution
is given by the perturbative result and an additional non-perturbative
contribution can be evaluated in a well-defined manner \cite{yaf}. Such a
split into a dominant perturbative contribution and a non-perturbative
remainder should also be possible for the free energy in the symmetric
phase. This would then justify the perturbative treatment of the first-order 
electroweak phase transition. For the pure gauge theory this problem has
been discussed in \cite{brat}. 

\begin{figure}[bh]
\begin{center}
\epsfig{file=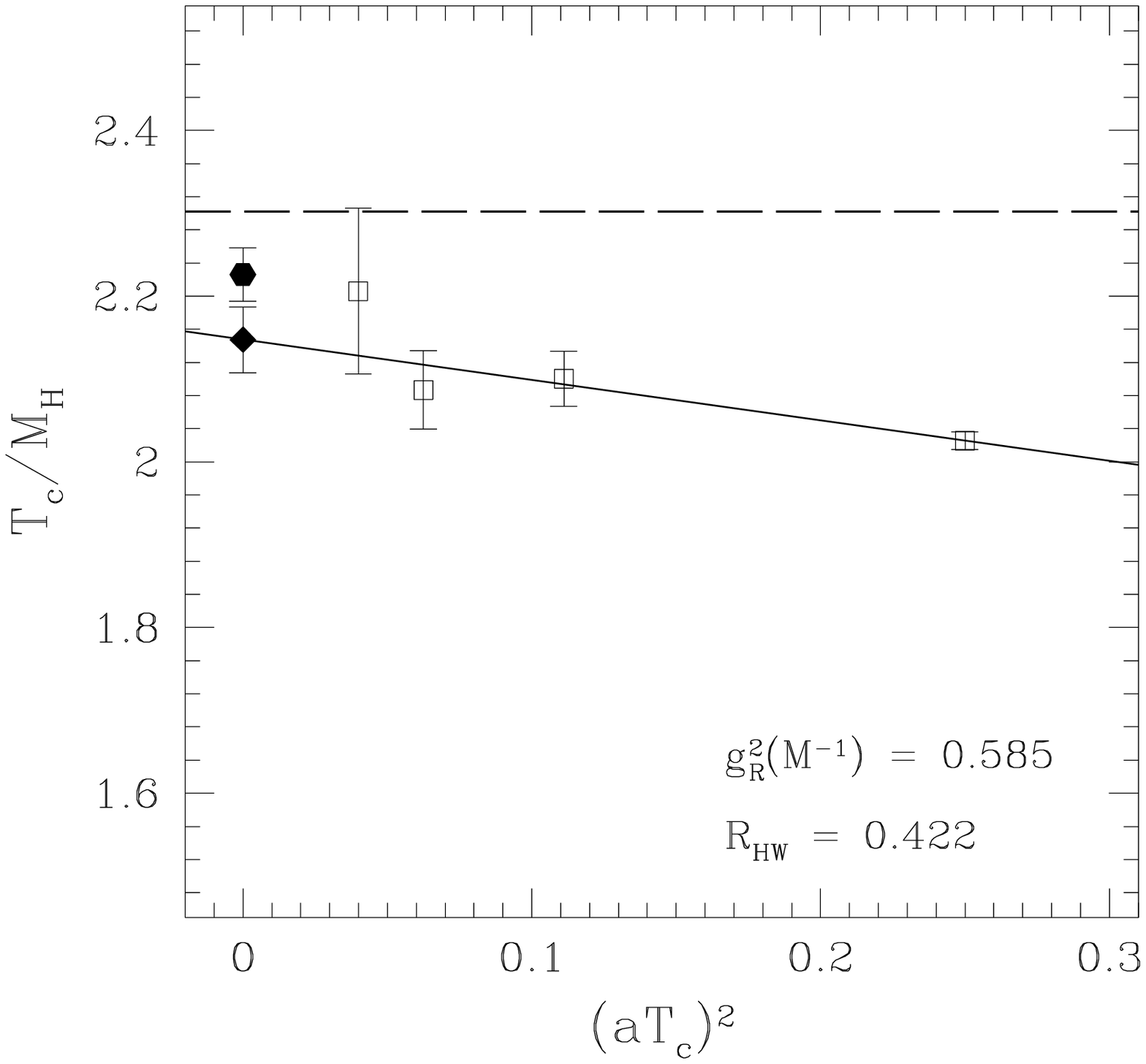,width=10cm}
\caption{Numerical results for the ratio of critical temperature and Higgs
boson mass versus $(aT_c)^2=L_t^{-2}$; the full line is the extrapolation
to the continuum limit \protect{\cite{csikor}}. The dashed horizontal line 
shows the prediction of two-loop perturbation theory \protect{\cite{bfh}}. 
The hexagon between continuum limit and perturbation theory represents the 
result of a three-dimensional simulation \protect{\cite{kaj}}, adjusted to the 
parameters of the 
four-dimensional calculation \protect{\cite{laine}}. }
\end{center}
\end{figure}

Perturbation theory in the symmetric phase fails with respect to
correlation lengths of gauge-invariant vector and scalar fields. Lattice 
simulations yield shorter correlation lengths for vector fields in the
symmetric phase than in the Higgs phase \cite{montvay, kaj, jan, phil, gurt}.
The contrary is true for the correlation length of the vector field evaluated
in a fixed gauge \cite{buphi, karsch} - a puzzle which still remains to be
resolved. The latter screening lengths appear to be related to the critical
Higgs mass $m_H \sim 80$ GeV where the first-order phase transition turns
into a smooth crossover \cite{buphi, kaj2,rum}.

Independent of the intriguing non-perturbative features of the symmetric
phase, which require further studies, it is now known that the necessary
condition (\ref{spcon}) for electroweak baryogenesis is not satisfied in
the standard model. For Higgs boson masses above the present experimental
bound of 58 GeV \cite{pdg} one has according to Fig.~6,
\beq
{\D v(T_c) \over T_c} < 0.7 \; ,
\eeq
which is much smaller than the lower bound (\ref{spcon}). For appropriate
choices of parameters it is possible to satisfy this bound in some
extensions of the standard model, but in these models it has not
been demonstrated that electroweak baryogenesis is indeed possible. This
strongly suggests that the baryon asymmetry has been generated in the
high-temperature phase of the standard model, as it is the case for instance
in conventional grand unified theories. 

\section{Baryogenesis via leptogenesis}

In the high temperature phase of the standard model the asymmetries of
baryon number and lepton number are proportional in thermal 
equilibrium (cf.~(\ref{bleq})),
\beqx 
\langle B \rangle_T \simeq C \langle B-L \rangle_T 
\simeq {C\over C-1} \langle L \rangle_T \, . 
\eeqx
In the standard model, as well as its
unified extension based on the group SU(5), $B-L$ is conserved. Hence,
no asymmetry in $B-L$ can be generated, and $\langle B \rangle_T$ vanishes.
Furthermore, as discussed above, baryogenesis at the electroweak phase
transition appears unlikely. As a consequence, the non-vanishing of
the baryon asymmetry is a strong argument for lepton number violation.
This is naturally realized by adding right-handed Majorana neutrinos
to the standard model. This extension of the standard model can be
embedded into grand unified theories with gauge groups containing
SO(10) \cite{fri}. Heavy right-handed Majorana neutrinos can also
explain the smallness of the light neutrino masses via the see-saw
mechanism \cite{seesaw}.

The connection between baryon number and lepton number at high temperatures
can be used to generate a baryon asymmetry. This was suggested by Fukugita and
Yanagida \cite{fy2}. 
The primordial lepton asymmetry is generated by the out-of-equilibrium
decay of heavy Majorana neutrinos in the standard manner. This
mechanism has subsequently been studied by several authors 
\cite{luty,etc,pluemi}, and it has been shown that the observed baryon 
asymmetry (cf.~\cite{kt}),
\beq
     Y_B={n_B\over s}=(0.6-1)\cdot10^{-10}\, ,
\eeq
can be obtained for a wide range of parameters. 

In unified theories with right-handed neutrinos $B-L$ is in general 
spontaneously broken. Unification also restricts the new parameters
which are introduced by adding right-handed neutrinos to the standard model.
In SO(10) unification it is natural to assume a similar pattern of mixings 
and masses for leptons and quarks. This ansatz, together with the requirement 
of baryogenesis, also restricts the scale of $B-L$ breaking. The following
discussion is closely related to \cite{buplu}.

The most general lagrangian for couplings and masses of charged leptons and 
neutrinos is given by 
\beq
  \cl_Y = -\overline{l_L}\,\tilde{\f}\,g_l\,e_R
          -\overline{l_L}\,\f\,g_{\n}\,\n_R
          -{1\over2}\,\overline{\n^C_R}\,M\,\n_R
          +\mbox{ h.c.}\;,
\eeq
where $l_L=\left(\n_L,e_L\right)$ is the left-handed lepton doublet
and $\f=(\vf^0,\vf^{-})$ is the standard model Higgs doublet. The
vacuum expectation value of the Higgs field $\VEV{\f}=v\ne0$ generates
Dirac masses $m_l$ and $m_D$ for charged leptons and neutrinos,
\beq
     m_l=g_lv \quad\mbox{and}\quad
     m_D=g_{\n}v\;,
\eeq
which are assumed to be much smaller than the Majorana masses $M$.
This yields light and heavy neutrinos
\beq
     \n\simeq K^{\dg}\n_L+\n_L^C K\quad,\quad
     N\simeq\n_R+\n_R^C\, ,
\eeq
with masses
\beq
     m_{\n}\simeq- K^{\dg}m_D{1\over M}m_D^T K^*\,
     \quad,\quad  m_N\simeq M\, ,
     \label{seesaw}
  \eeq
as mass eigenstates. Here $K$ is a unitary matrix which relates weak and
mass eigenstates. 
\begin{figure}[bh]
\begin{center}
\epsfig{file=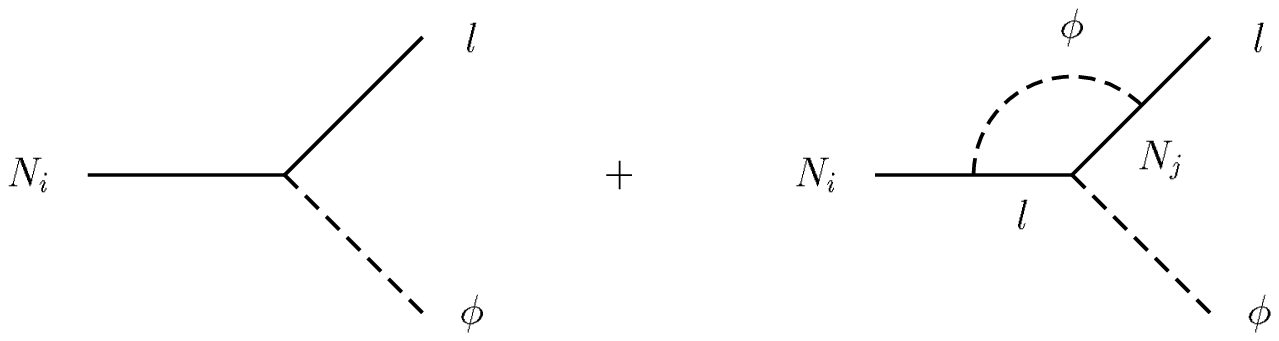,width=12cm}
\caption{Contributions to the decay of a heavy Majorana neutrino}
\end{center}
\end{figure}
Since the heavy neutrinos $N_i$ are Majorana fermions, 
their decay to lepton and Higgs scalar violates lepton number.
In the rest system the decay width of $N_i$ reads at tree level,
\beq
    \G_{Di}:=\G_{rs}\left(N^i\to\f^{\dg}+l\right)+
    \G_{rs}\left(N^i\to\f+\overline{l}\right)=
    {M_i\over8\p}{(m_D^{\dg}m_D)_{ii}\over v^2}\;.
    \label{decay}
\eeq
Interference between tree level and one-loop amplitudes (cf.~Fig.~8) yields the
$CP$ asymmetry \cite{pluemi}
\beqa
  &&\e_i={1\over8\pi v^2\left(m_D^{\dag}m_D\right)_{ii}}\sum\limits_j
  \mbox{Im}\left[\left(m_D^{\dag}m_D\right)_{ij}^2\right]\,f
  \left({M_j^2\over M_i^2}\right)\label{cpasymm}\\[1ex]
  &&\quad\mbox{with}\quad f(x)=\sqrt{x}\left[1-(1+x)\ln\left({1+x\over x}
  \right)\right]\;.\NO
\eeqa
The corresponding maximal $B-L$ asymmetry is $\e_i/g*$, where $g*$ is the 
number of relativistic degrees of freedom (cf.~\cite{kt}).

In order to generate the observed baryon asymmetry several conditions have
to be fulfilled. First, the $CP$ asymmetry $\e_i$ has to be large enough;
second, the out-of-equilibrium condition $\G_{Di} < \k H(T=M_i)$ for the
decaying heavy neutrino has to be fulfilled, where $H$ is the Hubble parameter;
third, the $L$ violating interactions have to be sufficiently weak in order
not to erase the generated lepton asymmetry. These conditions tend to favour
small masses for the light neutrinos and a large scale of $B-L$ breaking
\cite{etc2}. However, these constraints are not model independent. The
corresponding bounds on the light neutrino masses can be considerably
relaxed if appropriate chiral symmetries are effectively conserved in
some temperature range in the symmetric phase \cite{kainu}.

All these conditions are automatically taken into account if one integrates
the Boltzmann equations including all relevant interactions for the model
under consideration. The results discussed below are based on such an
analysis using the Boltzmann equations described in \cite{pluemi}.
All three heavy  neutrino families are taken into account as intermediate 
states whereas only the asymmetry generated by the lightest of the
right-handed neutrinos is relevant, since the asymmetries generated by the 
heavier neutrinos are washed out.\\

{\noindent\it Neutrino masses and mixings}\\

Let us now consider a similar pattern of mixings and
mass ratios for leptons and quarks, which is natural in SO(10)
unification.  Such an ansatz is most transparent in a basis where all
mass matrices are maximally diagonal. In addition to real mass
eigenvalues two mixing matrices appear. One can always choose a basis
for the lepton fields such that the mass matrices $m_l$ for the
charged leptons and $M$ for the heavy Majorana neutrinos $N_i$ are
diagonal with real and positive eigenvalues,
  \beq
  m_l=\left(\begin{array}{ccc}m_e&0&0\\0&m_{\m}&0\\0&0&m_{\t}
  \end{array}\right)\qquad
  M=\left(\begin{array}{ccc}M_1&0&0\\0&M_2&0\\0&0&M_3
  \end{array}\right)\;.
  \eeq
In this basis $m_D$ is a general complex matrix, which can 
be diagonalized by a biunitary transformation. Therefore, we can
write $m_D$ in the form
  \beq
  m_D=V\,\left(\begin{array}{ccc}
  m_1&0&0\\0&m_2&0\\0&0&m_3\end{array}\right)\,U^{\dag}\;,
  \eeq
where $V$ and $U$ are unitary matrices and the $m_i$ are real and
positive. In the absence of a Majorana mass term $V$ and $U$ would 
correspond to Kobayashi-Maskawa type mixing matrices of left- and 
right-handed charged currents, respectively.

According to Eqs.~(\ref{decay}) and (\ref{cpasymm}) the $CP$ asymmetry
is determined by the mixings and phases present in the product
$m_D^{\dg}m_D$, where the matrix $V$ drops out.  Therefore, to
leading order, the mixings and phases which are responsible for
baryogenesis are entirely determined by the matrix $U$.
Correspondingly, the mixing matrix $K$ in the leptonic charged
current, which determines $CP$ violation and mixings of the light
leptons, depends on mass ratios and mixing angles and phases of $U$
and $V$.  Hence, there is no direct connection between the $CP$
violation and generation mixing at high and low energies.

Consider now the mixing matrix $U$. One can factor out five phases,
which yields
  \beq
    U=\mbox{e}^{i\g}\,\mbox{e}^{i\l_3\a}\,\mbox{e}^{i\l_8\b}\,U_1\,
    \mbox{e}^{i\l_3\s}\,\mbox{e}^{i\l_8\t}\;,
  \eeq
where the $\l_i$ are the Gell-Mann matrices. The remaining matrix
$U_1$ depends on three mixing angles and one phase, like the 
Kobayashi-Maskawa matrix for quarks. In analogy to the quark mixing
matrix we choose the Wolfenstein parametrization \cite{wolfenstein} as
ansatz for $U_1$,
  \beq\label{mm}
    U_1=\left(\begin{array}{ccc}
    1-{\l^2\over2}  &      \l        & A\l^3(\r-i\h)\\[1ex]
        -\l         & 1-{\l^2\over2} & A\l^2 \\[1ex]
    A\l^3(1-\r-i\h) &    -A\l^2      &  1
    \end{array}\right)\;,
  \eeq
where $A$ and $|\r+i\h|$ are of order one, while the mixing
parameter $\l$ is assumed to be small. For the masses $m_i$ and
$M_i$ we assume a hierarchy like for up-type quarks,
  \beqa
  m_1=b\l^4m_3&\quad m_2=c\l^2m_3&\quad b,c=\co(1)\\[1ex]
  M_1=B\l^4M_3&\quad M_2=C\l^2M_3&\quad B,C=\co(1)\;.\label{Mmass}
  \eeqa
For the eigenvalues $m_i$ of the Dirac mass matrix this choice is
motivated by SO(10) unification. The masses $M_i$ cannot be
degenerate, because in this case there exists a basis for $\n_R$
such that $U = 1$, which implies that no baryon asymmetry is
generated. For simplicity the masses $M_i$ are assumed to
scale like the Dirac neutrino masses.

The light neutrino masses are given by the seesaw formula
(\ref{seesaw}). The matrix $K$, which diagonalises the neutrino mass
matrix, can be evaluated in powers of $\l$. A straightforward calculation
gives the following masses for the light neutrino mass eigenstates
  \beqa
     m_{\n_e}&=&{b^2\over\left|C+\mbox{e}^{4i\a}\;B\right|}\;\l^4
             \;m_{\n_{\t}}+\co\left(\l^6\right)\label{mne}\\[1ex]
     m_{\n_{\m}}&=&{c^2\left|C+\mbox{e}^{4i\a}\;B\right|\over BC}
             \;\l^2\;m_{\n_{\t}}+\co\left(\l^4\right)\label{mnm}\\[1ex]
     m_{\n_{\t}}&=&{m_3^2\over M_3}+\co\left(\l^4\right)\;.\label{mnt}
  \eeqa

The $CP$-asymmetry in the decay of the lightest right-handed
neutrino $N_1$ is easily obtained from Eqs.~(\ref{cpasymm}) and
(\ref{mm})-(\ref{Mmass}),  
  \beq
    \e_1=-\;{1\over16\p}\;{B\;A^2\over c^2+A^2\;|\r+i\h|^2}\;\l^4\;
    {m_3^2\over v^2}\;\mbox{Im}\left[(\r-i\h)^2
    \mbox{e}^{i2(\a+\sqrt{3}\b)}\right]
    \;+\;\co\left(\l^6\right)\;.
  \eeq
This yields for the magnitude of the $CP$ asymmetry,
  \beq\label{cpa}
    |\e_1| \leq {1\over16\p}\;{B\;A^2\;|\r+i\h|^2\over c^2+A^2\;|\r+i\h|^2}\;
    \l^4\;{m_3^2\over v^2}\;+\;\co\left(\l^6\right)\;.
  \eeq
How close the value of $|\e_1|$ is to this upper bound depends on the phases
$\a$, $\b$ and $\arg{(\r+i\h)}$. 
Since $\e_1\propto m_3^2/v^2$, one can already conclude that a large value
of the Yukawa coupling $m_3/v$ will be preferred by this mechanism of 
baryogenesis. This holds irrespective of the neutrino mixings.\\

{\noindent\it Numerical results \label{results}}\\

To obtain a numerical value for the produced baryon asymmetry, one has to
specify the free parameters in the ansatz (\ref{mm})-(\ref{Mmass}). 
In the following we will  use as a constraint the value for 
the $\n_{\m}$-mass which is preferred by the 
MSW explanation \cite{msw} of the solar neutrino deficit (cf.~\cite{kir}),
  \beq
    m_{\n_{\m}}\simeq 3\cdot10^{-3}\;\mbox{eV}\;. \label{msw}
  \eeq
A generic choice for the free parameters is to take all $\co(1)$
parameters equal to one and to fix $\l$ to a value which is
of the same order as the $\l$ parameter of the quark mixing matrix,
  \beq
    A=B=C=b=c=|\r+i\h|\simeq 1\; ,\qquad \l\simeq 0.1\;. \label{p1}
  \eeq
{}From Eqs.~(\ref{mne})-(\ref{mnt}), (\ref{msw}) and (\ref{p1}) one
now obtains,
\beq
  m_{\n_e}\simeq 8\cdot10^{-6}\;\mbox{eV}\; , \quad
  m_{\n_{\t}}\simeq 0.15\;\mbox{eV}\; .\label{m1}
\eeq
Finally, a second mass scale has to be specified. In unified theories based 
on SO$(10)$ the Dirac neutrino mass $m_3$ is naturally equal to the 
top-quark mass,
  \beq\label{3t}
     m_3=m_t\simeq 174\;\mbox{GeV}\;.
  \eeq
This determines the masses of the heavy Majorana neutrinos $N_i$,
  \beq
     M_3 \simeq 2\cdot10^{14}\;\mbox{GeV}\; ,\label{M3}
\eeq
and, consequently, $M_1\simeq 2\cdot10^{10}\;\mbox{GeV}$ and 
$M_2\simeq 2\cdot10^{12}\;\mbox{GeV}$. {}From Eq.~(\ref{cpa}) one obtains
the $CP$ asymmetry $|\e_1| \simeq 10^{-6}$, where we have assumed maximal 
phases. The solution of the set of Boltzmann equations discribed in 
\cite{pluemi} now yields the $B-L$ asymmetry (see Fig.~9a),
  \beq
     Y_{B-L} \simeq 3\cdot10^{-10}\; ,
  \eeq
which is indeed the correct order of magnitude. The precise value depends
on unknown phases.

  \begin{figure}
     \begin{minipage}[t]{7cm}
     \mbox{ }\hfill\hspace{1cm}(a)\hfill\mbox{ }
     \epsfig{file=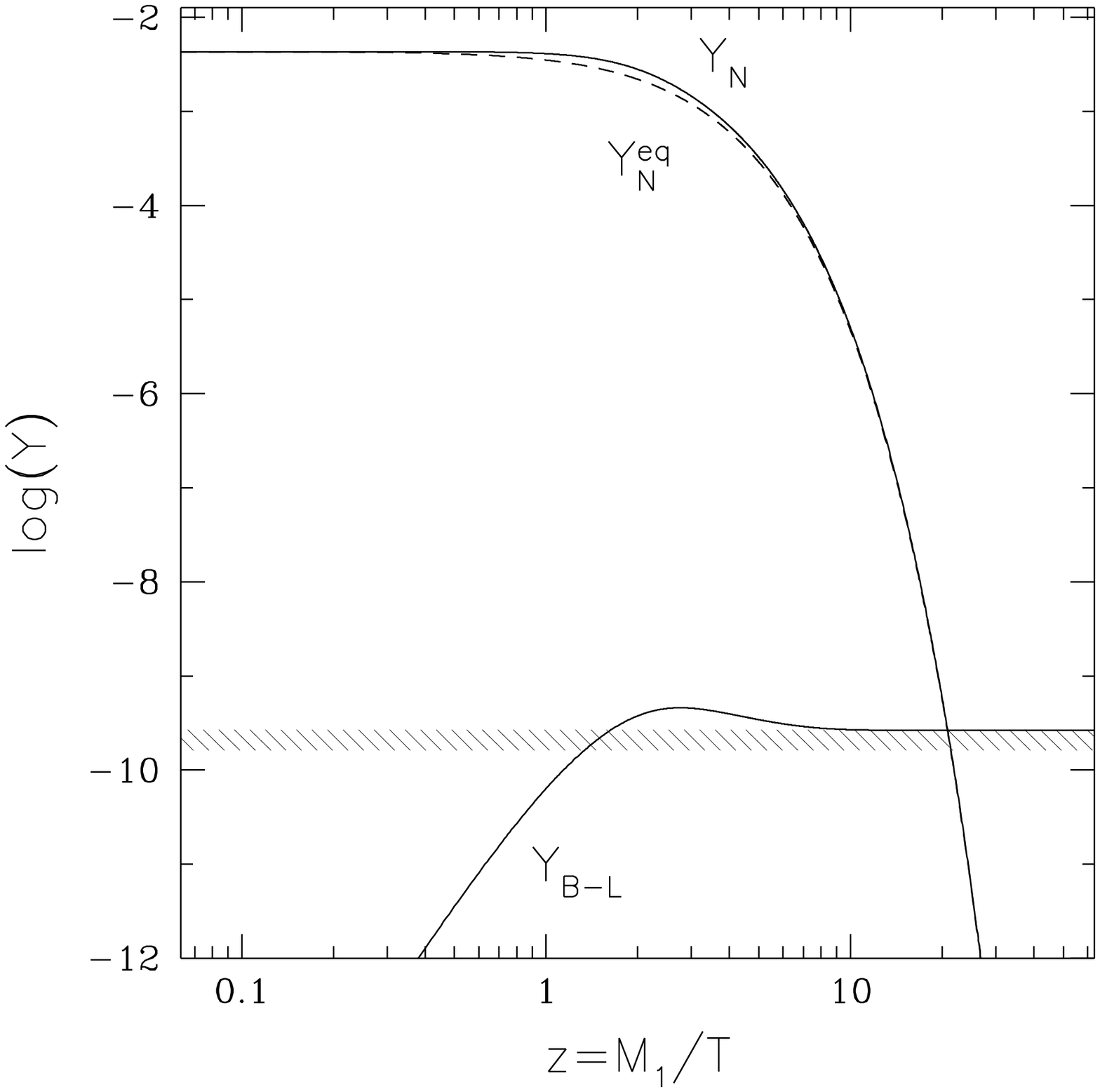,width=7cm}
     \end{minipage}
     \hspace{\fill}
     \begin{minipage}[t]{7cm}
     \mbox{ }\hfill\hspace{1cm}(b)\hfill\mbox{ }
     \epsfig{file=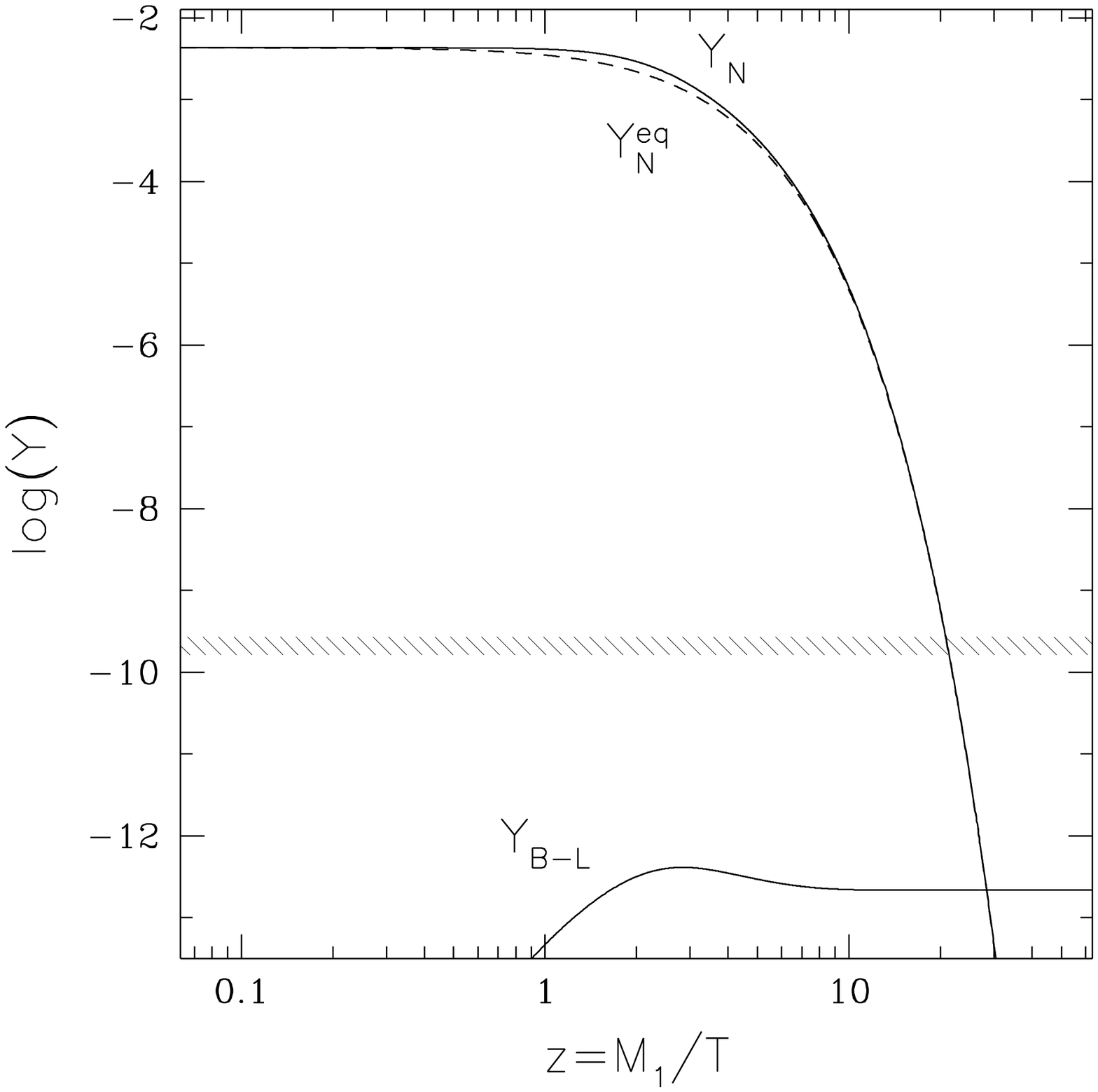,width=7cm}
     \end{minipage}  
     \caption{\it Time evolution of the neutrino number density and
     the $B-L$ asymmetry for $\l=0.1$ and for $m_3=m_t$ (a)
     or $m_3=m_b$ (b). The equilibrium distribution for
     $N_1$ is represented by a dashed line, while the hatched area
     shows the measured value for the asymmetry.}
  \end{figure}

The large mass $M_3$ of the heavy Majorana neutrino $N_3$ (cf.~(\ref{M3})), 
suggests that $B-L$ is already broken
at the unification scale $\Lambda_{\mbox{\scriptsize GUT}} \sim 10^{16}$
GeV, without any intermediate scale of symmetry breaking. This large
value of $M_3$ is a consequence of the choice  $m_3 \simeq m_t$. To test the 
sensitivity of the result for $Y_{B-L}$ on this assumption, consider the 
alternative choice, $m_3 = m_b \simeq 4.5$ GeV,
with all other parameters remaining unchanged. In this case one obtains
$M_3=10^{11}$ GeV and $|\e_1| = 5\cdot10^{-10}$ for the mass of $N_3$
and the $CP$ asymmetry, respectively. Since the maximal $B-L$ asymmetry is
$-\e_1/g*$ (cf.~\cite{kt}), it is clear that the generated asymmetry
will be too small. The solutions of the Boltzmann equations are shown
in Fig.~9b. The generated asymmetry, $Y_{B-L} \simeq 2\cdot10^{-13}$,
is too small by more than two orders of magnitude. We conclude that 
high values for both masses $m_3$ and $M_3$ are preferred, which is
natural in SO(10) unification.

Models for dark matter involving massive neutrinos favour a  
$\t$-neutrino mass $m_{\n_{\t}} \simeq 5\; \mbox{eV}$ \cite{raf},
which is significantly larger than the value given in (\ref{m1}).
Such a large value for the $\t$-neutrino mass can be accomodated within
the ansatz described in this section. However, it does not
correspond to the simplest choice of parameters and requires some
fine-tuning. For the mass of the heaviest Majorana neutrino one
obtains in this case $M_3 \simeq 6\cdot 10^{12}$ GeV.

Without an intermediate scale of symmetry breaking, the unification
of gauge couplings appears to require low-energy supersymmetry. This
provides further sources for generating a $B-L$ asymmetry \cite{cam}, 
whose size depends on additional assumptions. In this case, especially
constraints on the reheating temperature \cite{kt} and the 
possible role of preheating \cite{linde} require further studies.

\section{Conclusions}

The observation that baryon and lepton number violating processes are
in thermal equilibrium in the high-temperature phase of the standard
model, is of crucial importance for the theory of baryogenesis. 
In particular it implies that the presently observed cosmological
baryon asymmetry has been finally determined at the electroweak
phase transition. 

During the past three years a quantitative understanding of the first-order
electroweak phase transition for Higgs boson masses up to $m_H \sim 70$ GeV
has been achieved by means of analytical and numerical methods. The
transition to a crossover near $m_W \sim m_H$ and the full
understanding of the high-temperature phase still require
further work. However, already now we know that for Higgs boson masses
above the lower bound obtained at LEP, baryogenesis at the weak electroweak
transition is very unlikely.

Searching for alternatives to electroweak baryogenesis the connection
between baryon number and lepton number in the symmetric phase is again
crucial. It allows to generate the baryon asymmetry from a lepton
asymmetry, as suggested by Fukugita and Yanagida. Necessary ingredients
are right-handed neutrinos and Majorana masses, which appear naturally
in SO(10) unified theories. The example described in the previous section
demonstrates that this mechanism can explain the observed cosmological
baryon asymmetry without any fine-tuning of parameters. Further sources
for generating a lepton asymmetry exist in models with low-energy
supersymmetry. All this suggests that, hoping for further progress in
theory and new experimental results on neutrino properties, we can look
forward to an intriguing interplay between non-perturbative processes 
in the standard model, early universe cosmology and neutrino physics.
\\
\\ 
\\
{\bf\large Acknowledgements}\\
\\
It is a pleasure to thank Z.~Fodor, A.~Hebecker and M.~Pl\"umacher for
an enjoyable collaboration which led to the results described in this report.
I would also like to thank J.~Hein, K.~Jansen and I.~Montvay for their 
continuous help in understanding lattice results and for many valuable 
discussions. Last, but not least, I would like to thank the organizers of 
{\it Quarks '96} 
for a stimulating meeting and also for hospitality in Yaroslavl.

\pagebreak

\end{document}